\documentclass[pra,twocolumn,showpacs,superscriptaddress]{revtex4-1}
\usepackage{amsmath,amscd,amssymb,color}
\usepackage{graphicx,amsfonts,epsf}
\usepackage{epstopdf}
\usepackage{hyperref}
\usepackage{enumerate}
\usepackage{xcolor}
\newcommand{\ie}{\textit{i.e.~}}
\begin{document}
\title{Experimental Detection of Non-local Correlations 
using a Local Measurement-Based Hierarchy on an NMR Quantum Processor}
\author{Amandeep Singh}
\email{singh@sustech.edu.cn}
\affiliation{Shenzhen Institute for Quantum Science and
Engineering, and Department of Physics,
Southern University of Science and Technology, Shenzhen 518055, China}
\affiliation{Department of Physical Sciences, Indian
Institute of Science Education \& Research Mohali, Sector 81
SAS Nagar, Manauli PO 140306 Punjab India.}
\author{Dileep Singh}
\email{dileepsingh@iisermohali.ac.in}
\affiliation{Department of Physical Sciences, Indian
Institute of Science Education \& Research Mohali, Sector 81
SAS Nagar, Manauli PO 140306 Punjab India.}
\author{Vaishali Gulati}
\email{vaishali@iisermohali.ac.in}
\affiliation{Department of Physical Sciences, Indian
Institute of Science Education \& Research Mohali, Sector 81
SAS Nagar, Manauli PO 140306 Punjab India.}
\author{Arvind}
\email{arvind@iisermohali.ac.in}
\affiliation{Department of Physical Sciences, Indian
Institute of Science Education \& Research Mohali, Sector 81
SAS Nagar, Manauli PO 140306 Punjab India.}
\author{Kavita Dorai}
\email{kavita@iisermohali.ac.in}
\affiliation{Department of Physical Sciences, Indian
Institute of Science Education \& Research Mohali, Sector 81
SAS Nagar, Manauli PO 140306 Punjab India.}
\begin{abstract}
The non-local nature of the correlations possessed by
quantum systems may be revealed by experimental
demonstrations of the violation of Bell-type inequalities.
Recent work has placed bounds on the correlations that
quantum systems can possess in an actual experiment. These
bounds were limited to a composite quantum system comprising
of a few lower-dimensional subsystems. In a more general
approach, it has been shown that fewer body correlations can
reveal the non-local nature of the correlations arising from
a quantum mechanical description of nature. Such tests on
the correlations can be transformed to a semi-definite
program (SDP). This study reports the experimental
implementation of a local measurement-based hierarchy on the
nuclear magnetic resonance (NMR) hardware utilizing three
nuclear spins as qubits.  
The protocol has been experimentally tested on genuinely
entangled tripartite states such as W state, GHZ state and a
few graph states. In all the cases, the experimentally
measured correlations were used to formulate the SDP, using
linear constraints on the entries of the moment matrix. We
observed that for each genuinely entangled state, the SDP
failed to find a semi-definite positive moment matrix
consistent with the experimental data. This implies that the
observed correlations can not arise from local measurements
on a separable state and are hence non-local in nature,
and also confirms that the states being tested are indeed
entangled. 
\end{abstract} 
\maketitle 
\section{Introduction} 
It is well established that quantum computation has a
computational advantage over its classical counterpart and
the main resources utilized for quantum computation are
superposition, entanglement and other quantum features like
contextuality \cite{nielsen-book-02}. Entanglement plays a
key role in several quantum computational tasks, including
quantum teleportation \citep{bennett-prl-93}, quantum super
dense coding \citep{bennett-prl-92}, measurement-based
quantum computation \citep{briege-nature-09} and quantum key
distribution schemes \citep{ekert-prl-91,barrett-prl-05,
acin-njp-06}. Creating entangled states in an experiment and
certifying the presence of entanglement in such states is
hence of utmost importance \citep{guhne-pr-09,
horodecki-rmp-09} from a practical as well as the
foundational points of view.  Most of the known entanglement
detection schemes rely on experimental quantum state
reconstruction \citep{horodecki-rmp-09}, however it has been
shown that quantum state reconstruction is not
cost-effective with respect to experimental and
computational resources \citep{haffner-nature-05}.
Furthermore, the detection of entanglement of a known
quantum state is computationally a hard problem
\citep{brandao-prl-12} and scales exponentially with the
number of qubits. Methods to detect entanglement include the
use of violation of Bell-type inequalities
\citep{huber-pra-11, jungnitsch-prl-11}, entanglement
witnesses \citep{lewenstein-pra-00,guhne-jmo-03},
expectation values of the Pauli operators
\citep{zhao-pra-13, miranowicz-pra-13} as well as dynamical
learning techniques \citep{behrman-qic-08}. Although a
number of schemes exist for entanglement detection, most of
them lack generality and therefore more research is required
in this direction.

A promising direction of research on entanglement is to
experimentally observe the violation of Bell-type
inequalities \citep{bell-book, bell-ppf-64, chsh-prl-69},
and such inequalities have recently been proposed as a
general method for certifying the non-local nature of the
experimentally observed correlations in a device-independent
manner \citep{npa-prl-07, pironio-siam-10,flavio-prx-17}.
Consider a joint probability distribution $ P_{\alpha\beta}
$. The question addressed in Ref.\citep{npa-prl-07} is that
can there be a quantum description of $ P_{\alpha\beta} $
\ie can one have a quantum state $ \rho $, acting on the
joint Hilbert space $ \mathcal{H}^{}_A \otimes
\mathcal{H}^{}_B $, and the local measurement operators $
E_{\alpha}=\tilde{E}_{\alpha}\otimes I $ and $
E_{\beta}=I\otimes\tilde{E}_{\beta} $ such that
\begin{equation}\label{probabilitydistr}
P_{\alpha\beta}=Tr(E_{\alpha}^{}E_{\beta}^{}.\rho)
\end{equation}
where $ \tilde{E}_{\alpha} $ and $ \tilde{E}_{\beta} $ are
local projection operators. This can be used to design a
test for the detection of non-locality from the actual
probability distribution $ P_{\alpha\beta} $. In general,
one may need to search over all physical $ \rho $ and
projection operators $ E_{\mu} $ which makes the problem
computationally hard. A few attempts have been made to solve
this problem, including finding the maximum violation of the
Clauser Horne Shimony Holt (CHSH)
inequality~\citep{chsh-prl-69} by a quantum description
\citep{tsirelson-lpm-80}.  Notable work has been done by
Landau~\citep{landau-fp-88} and Wehner~\citep{wehner-pra-06}, 
where they have showed that the test
of whether the experimental correlations arise from quantum
mechanical description of nature or not, can be transformed
to a semi-definite program (SDP). Solving such an SDP can
reveal the local or non-local nature of the observed
correlations.

In this article we describe the details of the experimental
implementation of a local measurement-based hierarchy to
certify the non-local correlations arising from local
measurements~\citep{npa-prl-07,
pironio-siam-10,flavio-prx-17} on a system of three NMR qubits.
The motivation of the investigation is to experimentally
implement a non-local correlation and thereby devise an
entanglement detection  protocol which can readily be
generalized to higher numbers of qubits as well as to
multi-dimensional quantum systems. We experimentally
generate high fidelity genuinely entangled multipartite
states using three NMR qubits.  In order to experimentally
test the NPA hierarchy the desired correlators were measured
with high precision. In our experimental investigations,
these correlators are the three-qubit Pauli observables. The
expectation values of the Pauli observables were used to
formulate the moment matrix of SDP. Our earlier developed
techniques~\citep{singh-pra-16,singh-pra-18,singh-qip-18}
were utilized to find the expectation values of the desired
correlators, in a given state of the ensemble. SDP
optimization failure in finding a positive moment matrix is
an indication that the correlations encoded in the observed
correlators are non-local in nature and hence the state
under investigation is entangled. Experimental verification
was done by formulating the SDP by directly computing the
correlators from the tomographed state. Further, we note
here that the subsystems involved in our experiments reside
on the same molecule and therefore strictly speaking, it is
not possible to achieve a space-like separation between the
events occurring in the different subsystem spaces.
Therefore, the term ``local'' here pertains to subsystems
and non-local implies something that goes across subsystems.

The paper is organized as follows: Sec.-\ref{Theory}
introduces the main features of local measurement based
hierarchy and the formulation of the corresponding SDP,
while Sec.-\ref{m-NPA} outlines the modified hierarchy
obtained by introducing a relaxation due to commuting local
measurements \citep{npa-prl-07}. Sec.-\ref{tripartite-sdp}
details the SDP formulation, and the experimental
implementation using NMR. Sec.-\ref{remarks} contains a few
concluding remarks.
\section{Revisiting Local Measurement-Based Hierarchy}
\label{Theory}
To discuss the entanglement detection scenario, consider an
N-partite quantum state $\rho_N$, which is shared by N observers.
The joint probability distribution can be considered to
arise from the local measurements on a separable state $
\rho_N $. 
As the state $\rho_N$ is shared amongst $N$
parties, each of which can perform `$m$' different
measurements, each such measurement can result in `$d$'
different outcomes. Measurement by the $ i^{\mathrm{th}}$
party is represented by observables $ M_{x_i}^{a_i} $ with $ x_i\in
\lbrace 0,...,m-1 \rbrace $ being the measurement choice and
$ a_i\in \lbrace 0,...,d-1 \rbrace $ being the corresponding
outcome. By observing the statistics generated by measuring
all possible $ M_{x_i}^{a_i} $, one may write the empirical
values for the conditional probability distributions as
\begin{equation}\label{cond-Prob}
p(a_1,...,a_N\vert x_1,...,x_N)=Tr(M_{x_1}^{a_1}\otimes ...\otimes M_{x_N}^{a_N}\rho_N)
\end{equation}
The correlations observed by measuring $ M_{x_i}^{a_i} $
locally, get encoded in the conditional probability
distributions having the form of Eq.(\ref{cond-Prob}).
Similar expressions can be written for the reduced state
probability distributions which may arise from local
measurements on a reduced system.
\begin{equation}\label{cond-Red-Prob}
p(a_{i_1},...,a_{i_k}\vert x_{i_1},...,x_{i_k})=Tr(M_{x_{i_1}}^{a_{i_1}}\otimes...\otimes M_{x_{i_k}}^{a_{i_k}}\rho_{i_1.....i_k}) \nonumber
\end{equation}
where $ 0\leq i_1 <...<i_k<N $, \;$  1\leq k <N $ and $
\rho_{i_1.....i_k} $ is the reduced density matrix obtained
from $  \rho_N $. While dealing with dichotomic measurements
on qubits, it is useful to introduce the concept of
correlators and their expectation values as follows

\begin{eqnarray}\label{Expectation}
\langle M_{x_{i_1}}^{(i_1)}...M_{x_{i_k}}^{(i_k)} \rangle&=& \nonumber\\
\sum_{a_{i_1},...a_{i_k}}&(-1)&^{\sum_{l=1}^k a_{i_l}}p(a_{i_1},...a_{i_k}\vert x_{i_1},...x_{i_k})
\end{eqnarray}
The index $k$ here dictates the order of the correlator
while $0 \leq i_1 < ... < i_k < N$ with $x_{i_j}\in \{
0,m-1\}$ and $1\leq k \leq N$. One may note that for
dichotomic measurements it is convenient to introduce
$M_{x_i}^{(i)}=M_{x_i}^{(1)}-M_{x_i}^{(0)}$ and the $\langle
M_{x_{i_1}}^{(i_1)}...M_{x_{i_k}}^{(i_k)}
\rangle=Tr(M_{x_{i_1}}^{(i_1)}\otimes...\otimes
M_{x_{i_k}}^{(i_k)}\rho_{i_1,...,i_k})$. For $ k=2 $ the
correlator will be of second order having the form $\langle
M_{x_{i_1}}^{(i_1)} M_{x_{i_2}}^{(i_2)}\rangle$ while for $
k=N $ one can have the full body correlator. It will be seen
later that these correlators in the simplest case turn out
to be multi-qubit Pauli operators. For example, using
Eq.(\ref{Expectation}), for a two-qubit system the $\langle
M_{3}^{(1)}M_{3}^{(2)} \rangle=\langle
\sigma^{(1)}_z\otimes\sigma^{(2)}_z
\rangle=(-1)^{0+0}p(0,0\vert
\sigma^{(1)}_z,\sigma^{(2)}_z)+(-1)^{0+1}p(0,1\vert
\sigma^{(1)}_z,\sigma^{(2)}_z)+(-1)^{1+0}p(1,0\vert
\sigma^{(1)}_z,\sigma^{(2)}_z)+(-1)^{1+1}p(1,1\vert
\sigma^{(1)}_z,\sigma^{(2)}_z)$ which implies
that the expectation value of a correlator is the sum of the
products of various outcome probabilities with the
respective eigenvalues.

The method to detect entanglement, introduced in
Ref.\citep{flavio-prx-17}, assumes that:\\
\begin{itemize}
\item[(1)] Local measurements on a separable state $\rho_N$
can produce local correlations exhibiting local models.
\item[(2)] Local commuting measurements on a quantum state
can reveal all the local correlations.  
\item[(3)] A
positive moment matrix can be defined using correlations
produced by commuting local measurements, considering the
constraints due to commutation of measurements.
\end{itemize} 
The main assumption above can be explained as
follows. Consider that state $\rho_N$ is a separable state
and thus can be written as
$\rho_N=\sum_{\lambda}p_{\lambda}\otimes_{i}\rho_{\lambda}^{i}$.
One can write the conditional probabilities produced by
local measurements on such a separable state as
\begin{eqnarray}\label{separ} p(a_1,...,a_N\vert
x_1,...,x_N)&=&\sum_{\lambda}p_{\lambda}Tr(\otimes_i
M_{x_i}^{a_i}\otimes_i \rho_{\lambda}^{i})\nonumber\\
&=&\sum_{\lambda}p_{\lambda}\prod_{i}^{N}p(a_i\vert x_i,
\lambda) \end{eqnarray} As in Bell nonlocality, the
probability distributions exhibiting the above form are
local in nature and they do not violate any Bell-type
inequality. Conversely, if one can not write an observed
distribution in the above form, then the distribution is
non-local. Hence whenever the conditional probability
distributions given by Eq.(\ref{cond-Prob}) are non-local,
\ie do not exhibit the form of Eq.(\ref{separ}), the
state under consideration possesses non-local correlations
and is thus entangled.

In order to define the SDP for the
Navascu\'es-Pironio-Ac\'{\i}n (NPA) hierarchy based on local
measurements, consider the projectors $ E_{\nu} $ and $
E_{\mu} $,  corresponding to outcomes of a measurement $M$,
labeled by $\nu$ and  $\mu$. Here $M$ may or may not be a
projective measurement. These projectors satisfy the
following constraints:
\begin{enumerate}[(i)]
\item are orthogonal \ie $ E_{\nu}E_{\mu}=0$ for $ \nu $, $
\mu \in$ M, $ \mu\neq\nu $.
\item sum to identity \ie $ \sum_{\mu\in M} E_{\mu}=I $.
\item obey $ E_{\mu}^2=E_{\mu}^{\dagger}=E_{\mu} $.
\item obey the commutation rule (for projectors on
subsystems A and B) as $ [ E_{\alpha},E_{\beta}]=0 $.
\end{enumerate}

It was assumed in Ref.~\citep{npa-prl-07} that such a $ \rho
$ exists which satisfies Eq.(\ref{probabilitydistr}) and the
projector constraints. It was noted that by taking products
of projection operators $E_{\mu} $ and linear superpositions
of such products, one may define new operators which may
neither be projectors anymore nor be Hermitian. Let $
S=\lbrace S_1,S_2,....,S_n \rbrace$ be a set of $n$ such
operators. There exists an $ n \times n $ matrix associated
with every such set $S$ and defined as 
\begin{equation}\label{moment-matrix}
\Gamma_{ij}=Tr(S_i^{\dagger}S_j \rho)
\end{equation}
$\Gamma $ is Hermitian and satisfies
\begin{equation}\label{sdpCons1}
\sum_{i,j} c_{ij}S_i^{\dagger}S_j=0 \;\;\;  \Rightarrow \;\;\; \sum_{i,j} c_{ij}\Gamma_{ij}=0
\end{equation}
\begin{equation}\label{sdpCons2}
\sum_{i,j}
c_{ij}S_i^{\dagger}S_j=\sum_{\alpha,\beta}d_{\alpha\beta}E_{\alpha}E_{\beta}
\;  \Rightarrow \; \sum_{i,j}
c_{ij}\Gamma_{ij}=\sum_{\alpha,\beta}d_{\alpha\beta}P_{\alpha\beta}
\end{equation}
Further, it can be proved that $ \Gamma $ is positive
semi-definite \ie $ \Gamma \geq 0 $ \citep{npa-prl-07}.
Hence,
if a joint probability distribution $ P_{\alpha\beta} $ has
a quantum description \ie there exists a state $ \rho $ and
local measurement operators satisfying
Eq.(\ref{probabilitydistr}) and projector constraints
respectively, then finding such a state is equivalent to
finding the matrix $ \Gamma \geq 0 $ satisfying linear
constraints similar to Eq.~(\ref{sdpCons1}) and
Eq.~(\ref{sdpCons2}). This amounts to solving an SDP
problem.
\subsection{Modified NPA hierarchy}
\label{m-NPA}
Having discussed the main features of the NPA
hierarchy\citep{npa-prl-07}, we now turn to the method to
detect non-local correlations. Consider a set $ O=\lbrace
O_i \rbrace $ with $ 1 \leq i \leq k $ and $ O_i $ being
some product of the measurement operators $ \lbrace \rm
M_{x_i}^{a_i} \rbrace $ or their linear combinations. One
can associate a $k \times k$ matrix with $O$ defined by
Eq.(\ref{moment-matrix}) as $
\Gamma_{ij}=$Tr$(O_i^{\dagger}O_j.\rho_N ) $. For a given
choice of measurements on a separable state: (a) $ \Gamma $
will be a positive semi-definite matrix, (b) matrix elements
of $ \Gamma $ satisfy the linear constraints similar to
Eq.(\ref{sdpCons1})-(\ref{sdpCons2}), (c) some of the matrix
elements of $ \Gamma $ can be obtained by experimentally
measuring the probability distribution and (d) some of the $
\Gamma $ matrix entries correspond to unobservables, as
entries of the moment matrix are expectation values of
observables, which necessarily need to be represented by
Hermitian operators. However, in case the element of the moment
matrix arises from a non-Hermitian operator, the
respective expectation value is unobservable and thus enters
the moment matrix as a parameter to be optimized in SDP.

Keeping these facts in mind, one can design a hierarchy-based 
test to see if a given set of correlations can arise
from an actual quantum realization by performing local
measurements on a separable state. One can define a set $
O_{\nu} $ consisting of products of `$ \nu $' local
measurement operators or linear superpositions of such
products. Once $ O_{\nu} $ is defined, one can look for
associated $ \Gamma \geq 0 $ satisfying constraints similar
to Eq.(\ref{sdpCons1})-(\ref{sdpCons2}) to see if a given
set of correlations can arise from actual local measurements
on a separable state. If no solution is obtained to such an
SDP, then this would imply that the given set of correlations
cannot arise by local measurements on a separable quantum
state and hence the correlations are non-local. One can
always find a stricter set of constraints by increasing the
value of $ \nu $ \ie testing the nature of correlations at
the next level of the hierarchy.

In the experimental demonstration, as suggested in
Ref.\citep{flavio-prx-17}, a set of commutating
measurements have been used to formulate the SDP \ie an
additional constraint is introduced on the entries of $
\Gamma $ such that local measurements also commutate. This
additional constraint considerably reduces the original
computationally-hard problem \citep{flavio-prx-17}. All the
ideas developed till now can be understood with an example.
Consider $N=2$, two dichotomic measurements per party at the
hierarchy level $\nu =2 $. Let the measurements be labeled
as $ A_x $ and $ B_y$ with $ x,y\in[0,1] $. Set of operators
is $ O_2=\lbrace I, A_0, A_1, B_0, B_1, A_0A_1, A_0B_0,
A_0B_1, A_1B_0, A_1B_1,B_0B_1 \rbrace $. One can write the
corresponding moment matrix $\Gamma $  as
\citep{flavio-prx-17}

\begin{widetext}
\begin{footnotesize}
\begin{equation}
\label{2QGamma}
\hspace{-2cm}
 \Gamma=\left(
\begin{array}{lllllllllll}
1 & \textcolor{red}{\langle A_0 \rangle} & \textcolor{black}{\langle A_1 \rangle} & \textcolor{blue}{\langle B_0 \rangle} & \textcolor{black}{\langle B_1 \rangle} & \color[rgb]{1,0.84,0}{v_1} & \color[rgb]{0,0.58,0}{\langle A_0 B_0 \rangle} & \textcolor{purple}{\langle A_0 B_1 \rangle} & \textcolor{orange}{\langle A_1 B_0 \rangle} & \textcolor{black}{\langle A_1 B_1 \rangle} & \textcolor[rgb]{0.49,0.62,0.75}{v_2} \\

 \textcolor{red}{\langle A_0 \rangle} & 1 & \color[rgb]{1,0.84,0}{v_1} & \color[rgb]{0,0.58,0}{\langle A_0 B_0 \rangle} & \textcolor{purple}{\langle A_0 B_1 \rangle} & \textcolor{black}{\langle A_1 \rangle} & \textcolor{blue}{\langle B_0 \rangle} & \textcolor{black}{\langle B_1 \rangle} & \color[rgb]{0.55,0,0}{v_3} & \textcolor{green}{v_4} & \textcolor{magenta}{v_5}\\

 \textcolor{black}{\langle A_1 \rangle} & \color[rgb]{1,0.84,0}{v_1^{\ast}} & 1 & \textcolor{orange}{\langle A_1 B_0 \rangle} & \textcolor{black}{\langle A_1 B_1 \rangle} & \textcolor{red}{v_6} & \color[rgb]{0.55,0,0}{v_3^{\ast}} & \textcolor{green}{v_4^{\ast}} & \textcolor{blue}{\langle B_0 \rangle} & \textcolor{black}{\langle B_1 \rangle} & \color[rgb]{0.55,0.55,0}{v_7}\\

\textcolor{blue}{\langle B_0 \rangle} & \color[rgb]{0,0.58,0}{\langle A_0 B_0 \rangle} & \textcolor{orange}{\langle A_1 B_0 \rangle} & 1 & \textcolor[rgb]{0.49,0.62,0.75}{v_2} & \color[rgb]{0.55,0,0}{v_3} & \textcolor{red}{\langle A_0 \rangle} & \textcolor{magenta}{v_5} & \textcolor{black}{\langle A_1 \rangle} & \color[rgb]{0.55,0.55,0}{v_7} & \textcolor{black}{\langle B_1 \rangle}\\

\textcolor{black}{\langle B_1 \rangle} & \textcolor{purple}{\langle A_0 B_1 \rangle} & \textcolor{black}{\langle A_1 B_1 \rangle} & \textcolor[rgb]{0.49,0.62,0.75}{v_2^{\ast}} & 1 & \textcolor{green}{v_4} & \textcolor{magenta}{v_5^{\ast}} & \textcolor{red}{\langle A_0 \rangle} & \color[rgb]{0.55,0.55,0}{v_7^{\ast}} & \textcolor{black}{\langle A_1 \rangle} & \textcolor{blue}{v_8}\\

\color[rgb]{1,0.84,0}{v_1^{\ast}} & \textcolor{black}{\langle A_1 \rangle} & \textcolor{red}{v_6^{\ast}} & \color[rgb]{0.55,0,0}{v_3^{\ast}} & \textcolor{green}{v_4^{\ast}} & 1 & \textcolor{orange}{\langle A_1 B_0 \rangle} & \textcolor{black}{\langle A_1 B_1 \rangle} & \textcolor[rgb]{0.24,0.7,0.44}{v_9} & \textcolor{purple}{v_{10}} & \color[rgb]{0,1,1}{v_{11}} \\

\color[rgb]{0,0.58,0}{\langle A_0 B_0 \rangle} & \textcolor{blue}{\langle B_0 \rangle} & \color[rgb]{0.55,0,0}{v_3} & \textcolor{red}{\langle A_0 \rangle} & \textcolor{magenta}{v_5} & \textcolor{orange}{\langle A_1 B_0 \rangle} & 1 & \textcolor[rgb]{0.49,0.62,0.75}{v_2} & \color[rgb]{1,0.84,0}{v_1} & \color[rgb]{0,1,1}{v_{12}} & \textcolor{purple}{\langle A_0 B_1 \rangle} \\

\textcolor{purple}{\langle A_0 B_1 \rangle} & \textcolor{black}{\langle B_1 \rangle} & \textcolor{green}{v_4} & \textcolor{magenta}{v_5^{\ast}} & \textcolor{red}{\langle A_0 \rangle} & \textcolor{black}{\langle A_1 B_1 \rangle} & \textcolor[rgb]{0.49,0.62,0.75}{v_2^{}\ast} & 1 & \color[rgb]{0,1,1}{v_{13}} & \color[rgb]{1,0.84,0}{v_1} & \color[rgb]{0,0.58,0}{v_{14}} \\

\textcolor{orange}{\langle A_1 B_0 \rangle} & \color[rgb]{0.55,0,0}{v_3^{\ast}} & \textcolor{blue}{\langle B_0 \rangle} & \textcolor{black}{\langle A_1 \rangle} & \color[rgb]{0.55,0.55,0}{v_7} & \textcolor[rgb]{0.24,0.7,0.44}{v_9^{\ast}}  & \color[rgb]{1,0.84,0}{v_1^{\ast}} & \color[rgb]{0,1,1}{v_{13}^{\ast}} & 1 & \textcolor[rgb]{0.49,0.62,0.75}{v_2} & \textcolor{black}{\langle A_1 B_1 \rangle} \\

\textcolor{black}{\langle A_1 B_1 \rangle} & \textcolor{green}{v_4^{\ast}} & \textcolor{black}{\langle B_1 \rangle} & \color[rgb]{0.55,0.55,0}{v_7^{\ast}} & \textcolor{black}{\langle A_1 \rangle} & \textcolor{purple}{v_{10}^{\ast}} & \color[rgb]{0,1,1}{v_{12}^{\ast}} &\color[rgb]{1,0.84,0}{v_1^{\ast}} & \textcolor[rgb]{0.49,0.62,0.75}{v_2^{}\ast} & 1 &  \textcolor{orange}{v_{15}}\\

\textcolor[rgb]{0.49,0.62,0.75}{v_2^{\ast}}& \textcolor{magenta}{v_5^{}\ast} & \color[rgb]{0.55,0.55,0}{v_7^{\ast}} & \textcolor{black}{\langle B_1 \rangle} & \textcolor{blue}{v_8^{\ast}} &  \color[rgb]{0,1,1}{v_{11}^{\ast}} & \textcolor{purple}{\langle A_0 B_1 \rangle} & \color[rgb]{0,0.58,0}{v_{14}^{\ast}} & \textcolor{black}{\langle A_1 B_1 \rangle} & \textcolor{orange}{v_{15}^{\ast}} & 1
\end{array}
\right)
\end{equation}
\end{footnotesize}
\end{widetext}
\noindent while following are the unassigned variables
\begin{flushleft}
\begin{footnotesize}
$
\begin{array} {lll}
v_1= \langle A_0A_1 \rangle, & v_2=\langle B_0B_1 \rangle,& v_3=\langle A_0A_1B_0 \rangle,\\

v_4=\langle A_0A_1B_1 \rangle, & v_5= \langle A_0B_0B_1 \rangle, & v_6=\langle A_1A_0A_1 \rangle, \\

v_7=\langle A_1B_0B_1 \rangle,& v_8=\langle B_1B_0B_1 \rangle, & v_9= \langle A_1A_0A_1B_0 \rangle, \\

 v_{10}=\langle A_1A_0A_1B_1 \rangle,& v_{11}=\langle A_1A_0B_0B_1 \rangle,& v_{12}=\langle A_0A_1B_0B_1 \rangle, \\

v_{13}= \langle A_0A_1B_1B_0 \rangle, & v_{14}=\langle A_0B_1B_0B_1 \rangle, & v_{15}=\langle A_1B_1B_0B_1 \rangle. 
\end{array}
$
\end{footnotesize}
\end{flushleft}
We note that by introducing local measurements commutativity \ie
$ [A_0,A_1]=[B_0,B_1]=0  $ the matrix elements, of the $ \Gamma $ matrix given
by Eq.(\ref{2QGamma}) were simplified. Specifically, the following reduction in
the number of variables can be noticed : $v_i=v_i^{\ast}$ for $i\in [1,15]$ and
$v_6=\langle A_0 \rangle$, $v_8=\langle B_0 \rangle$, $v_9=v_{14}=\langle
A_0B_0 \rangle$, $v_{10}=\langle A_0B_1 \rangle$,  $v_{15}=\langle A_1B_0
\rangle$ and also $v_{11}=v_{12}=v_{13}$. For a visual representation, the
variables that become identical because of the commutativity constraints are
represented by the same color in Eq.(\ref{2QGamma}) \citep{flavio-prx-17}. The
generated SDP will check if the set of observed correlations $  \lbrace
\langle A_x \rangle, \langle B_y \rangle, \langle A_xB_y \rangle \rbrace $ are
local. This can be achieved by substituting the experimental values of the
correlators in the $ \Gamma $ matrix and leaving the unobservables as variables.
The SDP will optimize over such variables 
to see if a given set of correlations are
local or non-local. It has been shown \citep{navascues-njp-08, pironio-siam-10}
that this method converges \ie if a given set of correlations are non-local
then the SDP will fail at a finite number of steps $\nu $.
\section{Detection of Tripartite Non-Local Correlations}
\label{tripartite-sdp}
A three-qubit system was used 
to experimentally demonstrate the detection of correlations
which can not arise from local measurements on a separable
state. 
It has been shown
\cite{dur-pra-00} that a genuine three-qubit system can be
entangled in two inequivalent ways. The CHSH scenario
\citep{chsh-prl-69} deals with the (2,2,2) case \ie $N=2$, $m=2$
and $d=2$. Any correlation violating the CHSH inequality
exhibits non-local nature in the sense that, in principle
one cannot write a local hidden variable theory which can
reproduce the observed statistics. In the current
experimental study, the scenarios are (3,2,2) and (3,3,2)
\ie three parties with two (or three) dichotomic observables
per party. The measurements of three parties are labeled as
$ A_x $ , $ B_y $  and $ C_z $ respectively with measurement
labels $ x,y,z \in [0,1] $ in the (3,2,2) scenario while $ x,y,z
\in [0,1,2] $ in the (3,3,2) scenario. One can construct set $
O_2 $ for three parties, the way it was done in the previous
section for $ N=2 $. As detailed in Ref.
\citep{flavio-prx-17}, to detect non-local correlations
arising from the W state, one needs to perform local
measurements $ M_0^{(i)}=\sigma_x $ and $ M_1^{(i)}=\sigma_z
$ for all three parties for the observables entering the
moment matrix associated with $ O_2 $ defined above. Here $
\sigma_{x/y/z} $ are the spin-1/2 Pauli operators. For GHZ
type states, the measurements to generate the statistics
were chosen as $ M_0^{(i)}=\sigma_x $ and 
$ M_1^{(i)}=\frac{1}{\sqrt{2}}(\sigma_z + \sigma_x)$. A full
body correlator is also introduced while detecting non-local
correlations generated by the GHZ state as such states are
not suitable for detection of non-local correlation using
fewer body correlators \citep{flavio-prx-17}. Further, for
graph states $ M_0^{(i)}=\sigma_x $, $ M_1^{(i)}=\sigma_z $
and $ M_2^{(i)}=\frac{1}{\sqrt{2}}(\sigma_z + \sigma_x) $ were
chosen as measurements.

\subsection{NMR implementation of the detection scheme}
\label{NMR Implementation}
\begin{figure}
\begin{flushleft}
\includegraphics[angle=0,scale=1.0]{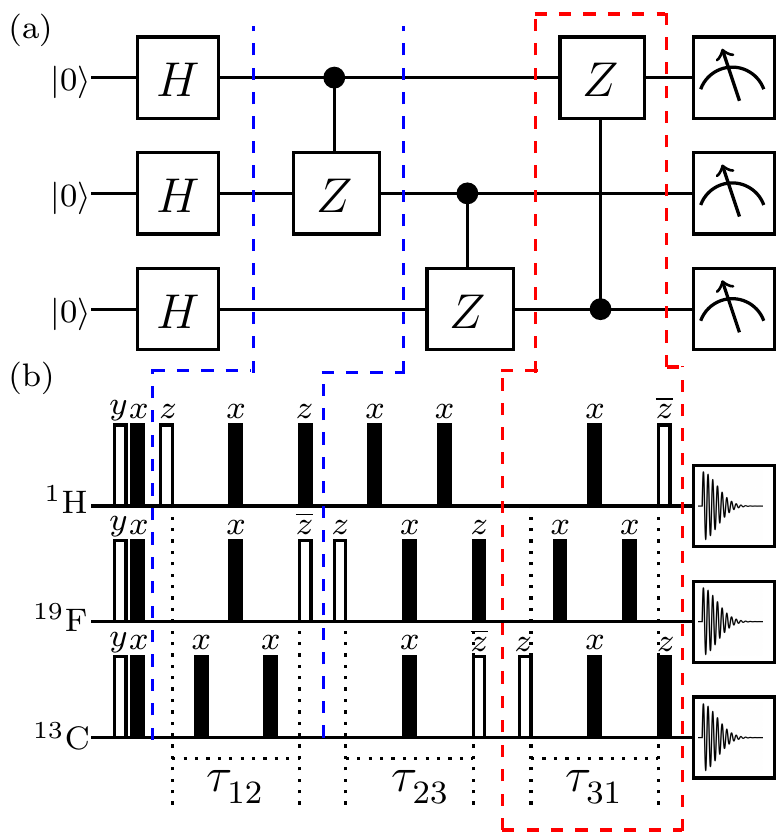}
\end{flushleft}
\caption{(a) Quantum circuit to prepare linear and loop
graph states on a three-qubit quantum processor. The third
controlled-Z gate, enclosed in red dashed line, acts only if
the circuit is being used to prepare the loop graph state.
(b) The NMR pulse sequence for the quantum circuit shown in
(a). The unfilled rectangles denote $\frac{\pi}{2}$
spin-selective RF pulses while the filled black rectangles
denote $\pi$ RF pulses. The phase of each pulse is written
above the respective pulse and a bar over the phase
represents negative phase. Delays are given by
$\tau^{}_{ij}=1/(8 J_{ij})$; $i,j$ are the qubit labels and
$J$ denotes the scalar coupling constant.}
\label{graph_state_ckt+seq} 
\end{figure}

A system of three NMR qubits was chosen to experimentally
demonstrate the detection of non-local correlations. The
system was initialized in either of the W, GHZ, linear or loop graph
states (see Ref.\citep{hein-pra-04} for the definition of
graph states). The quantum circuits as well as NMR pulse sequence
to prepare W and GHZ states are detailed in
Ref.\citep{singh-pra-16,singh-pra-18,singh-qip-18}. A
compact quantum circuit to prepare the linear as well as
loop graph state is shown in Fig.\ref{graph_state_ckt+seq}.
The third controlled-Z gate in
Fig.\ref{graph_state_ckt+seq} (enclosed by a red dotted box)
does not act while preparing the linear graph state.
The protocol to test the non-locality present in the
experimentally observed expectation values of the
correlators, for a given state, is as follows:
\begin{itemize}
\item The system was initialized in one of the genuine
tripartite pure states, either W, GHZ or graph states.
\item It was assumed that the correlations observed from
local measurements on such states will fail the SDP,
formulated in Sec.-\ref{Theory}, at the second level of the
modified NPA hierarchy.
\item At the second level ($ \nu=2 $) of the hierarchy, the
expectation values of all the correlators were measured
experimentally in the state under investigation.
\item Once all the observables of the moment matrix $ \Gamma
$ Eq.(\ref{moment-matrix}) were measured experimentally,
they were fed into the matrix $ \Gamma $. The remaining
unobservable entries of the moment matrix were left as
variables to be optimized via SDP, to achieve $ \Gamma \geq
0 $ under linear constraints similar to
Eq.(\ref{sdpCons1})-(\ref{sdpCons2}) as well as the
commutativity relaxation constraints \ie $
[A_0,A_1]=[B_0,B_1]=[C_0,C_1]=0 $ in (3,2,2) scenario.  
\item The above formulated SDP was solved by modifying the
codes, available at \cite{NPACode-github}, for the (3,2,2)
or (3,3,2) scenario.
\end{itemize} 

\subsection{NMR experimental set-up and system initialization}
For the experimental realization,$ ^{13} $C-labeled
diethylflouromalonate sample dissolved in acetone-D6 
is used, with  three spin-$\frac{1}{2}$ nuclei
\ie $ ^1 $H, $ ^{19} $F and $ ^{13} $C encoding the qubit 1,
qubit 2 and qubit 3, respectively. In such a scenario, the free Hamiltonian of the three-qubit NMR system in the rotating frame is given by
\citep{ernst-book-90}

\begin{equation}\label{NMR-Hamiltonian}
H=-\sum_{i=1}^3 \omega_i I_{iz}+2\pi\sum_{i,j=1}^3 J_{ij}I_{iz}I_{jz}
\end{equation}
where indices $i,\; j $= 1, 2 or 3 represent the qubit
number, $ \omega_i $ is the respective chemical shift, $
I_{iz} $ being the $z$-component of spin angular momentum
and $ J_{ij} $ is the scalar coupling constant. The experimental chemical shifts (in frequency units) were $\rm \nu_H$=3332.77 Hz,  $\rm \nu_F$=-110997.38 Hz and $\rm \nu_C$=12890.09 Hz. The longitudinal relaxation times were $\rm T^H_1$= 3.0$\pm$0.4 s, $\rm T^F_1$=3.3$\pm$0.2 s and  
$\rm T^C_1$=3.2 $\pm$ 0.4 s while the transverse relaxation times were measured as $\rm T^H_2$= 1.4$\pm$0.3 s, $\rm T^F_2$=1.3$\pm$0.2 s and  
$\rm T^C_2$=1.2 $\pm$ 0.3 s. Also the scalar coupling $\rm J_{HF}$=47.5 Hz, $\rm J_{FC}$=-191.5 Hz and $\rm J_{CH}$=161.6 Hz. Molecular structure and the representative NMR spectra of diethylflouromalonate can be found in Ref. \citep{singh-pra-18}. The system was
initialized in the pseudopure state (PPS) $ \vert 000
\rangle $ using the spatial averaging technique
\citep{cory-physD-98, mitra-jmr-07}

\begin{equation}
\rho^{}_{_{\mathrm{PPS}}}=\frac{(1-\epsilon )}{2^3}\mathbb{I}_8+\epsilon\vert 000 \rangle\langle 000 \vert \nonumber
\end{equation} 
where $ \epsilon\sim 10^{-5} $ is the room temperature
thermal magnetization and $ \mathbb{I}_8 $ is 8$ \times $8
identity matrix. The state fidelity of the experimentally prepared PPS was computed to be 0.96$ \pm $0.01 and was computed using the fidelity
measure \citep{uhlmann-rpmp-76, jozsa-jmo-94}
\begin{equation}
\rm F=\left[Tr\left(\sqrt{\sqrt{\rho_{theory}}\rho_{exptl}\sqrt{\rho_{theory}}}\right)\right]^2
\end{equation}
where $ \rm \rho_{theory} $ and $ \rm \rho_{exptl} $ are the
theoretically expected and experimentally observed density
operators. Full quantum state tomography
\citep{leskowitz-pra-04} was performed for the experimental
reconstruction of the density matrix using a set of seven
preparatory RF pulses \ie $ \lbrace III, XXX, IIY, XYX, YII,
XXY, IYY \rbrace  $. Here $I$ implies `no pulse' while
$X(Y)$ represents a $ \frac{\pi}{2} $  local rotation with
phase $x(y)$. In NMR such local unitary operations can be
achieved using highly precisely calibrated spin selective
radio frequency (RF) pulses. Non-local unitary operations
can be achieved by letting the system evolve freely under
system Hamiltonian (Eq.(\ref{NMR-Hamiltonian})) and the
desired scalar coupling $ J_{ij} $ by means of $ \pi
$-refocusing RF pulses suitably embedded in the free
evolution periods. Fidelities  of the experimentally
prepared W, GHZ, linear and loop graph states were 0.95$ \pm
$0.02, 0.96$ \pm $0.01, 0.95$ \pm $0.01 and 0.94$ \pm $0.02
respectively.

\subsection{Non-locality detection by experimentally
measuring the moments/correlators}
At the second level of the modified NPA hierarchy in the
(3,2,2) scenario, the set
\linebreak $ O_2=\lbrace \mathbb{I}_8,A_0, A_1, B_0, B_1,
C_0, C_1, A_0A_1, A_0B_0, A_0B_1, A_0C_0,$ $A_0C_1,A_1B_0,
A_1B_1, A_1C_0, A_1C_1, B_0B_1, B_0C_0, B_0C_1, B_1C_0,$
$B_1C_1, C_0C_1 \rbrace $. The moment matrix in this case is
a 22 $ \times $ 22 matrix with all diagonal entries as 1.
Further, the matrix has 26 observable moments while rest of
the moments enter the moment matrix as unobservables and
were left as variables to be optimized in SDP as detailed in
Sec.-\ref{NMR Implementation}.

\begin{figure}
\begin{flushleft}
\includegraphics[scale=0.8]{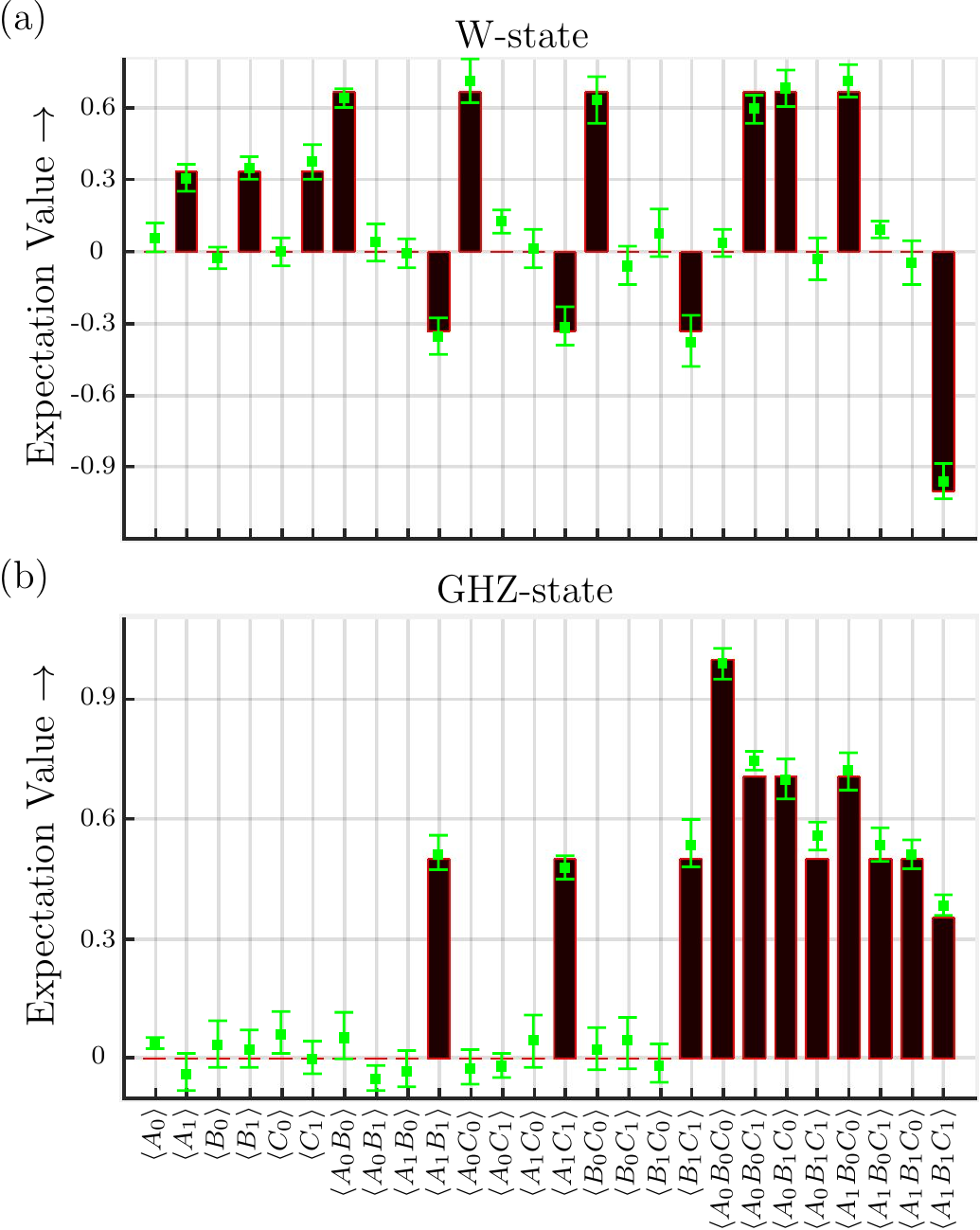}
\end{flushleft}
\caption{Bar plots for the observable moments of the moment matrix $\Gamma$ for (a) W and (b) GHZ states. Bars represent theoretically expected values while green squares are the experimentally observed values.}
\label{gammaplots}
\end{figure}
\begin{figure}[b]
\begin{flushleft}
\includegraphics[scale=0.8]{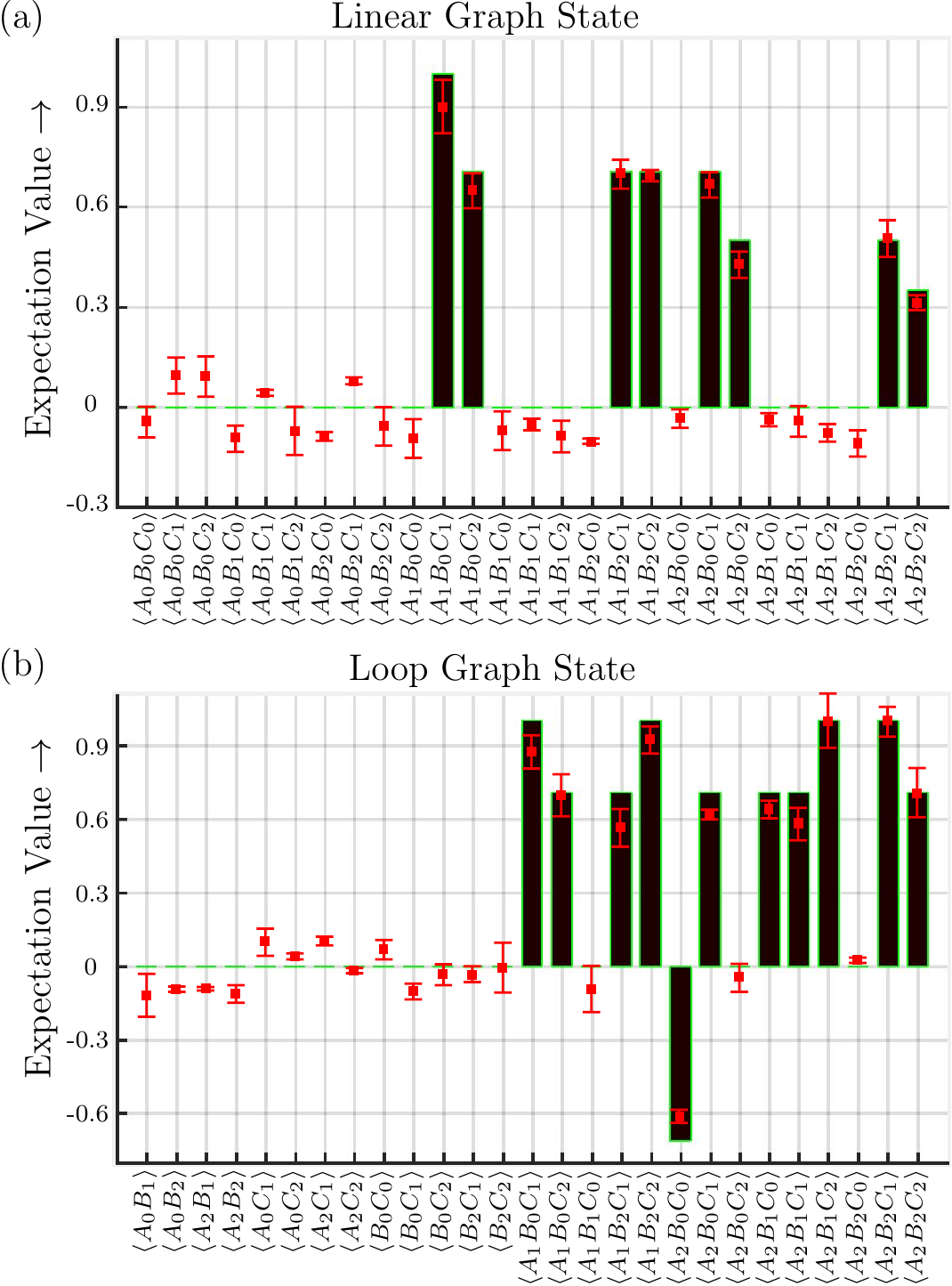}
\end{flushleft}
\caption{Bar plots for the observable moments of the moment matrix $\Gamma$ for (a) Linear and (b) Loop graph states. Bars represent theoretically expected values while green squares are the experimentally observed values.}
\label{gammaplots-grpahstates}
\end{figure}

As an example, the moment/correlator $ \Gamma_{4,12} $ in the case of W state,
is an observable $ \sigma_{1x} \sigma_{2x} \sigma_{3z} $
while the moment/correlator $ \Gamma_{1,18} $ is $
-i\sigma_{2y} $ which is not an observable and enters the
moment matrix as an SDP variable. Similarly one can compute
the parameters for (3,3,2) scenario. The next task was to
find the expectation values of the correlators in the state
under investigation. In NMR experiments the observed signal
is proportional to the $z$-magnetization of the ensemble
which indeed is proportional to the expectation value of the
Pauli $z$-spin angular momentum operator in the given state.
Hence the direct observable in a typical NMR experiment is
the Pauli $z$-operator expectation values of the nuclear
spins. We have previously developed schemes to find the
expectation values of any desired Pauli operators in the
given state \citep{singh-pra-16,singh-pra-18,singh-qip-18}.
This was achieved by mapping the state $ \rho\rightarrow
\rho_l=U_l.\rho.U_{l}^{\dagger} $ followed by
$z$-magnetization measurement. The explicit forms of the
unitary operators $ U_l $, as well as quantum circuits and
NMR pulse sequences, for two and three-qubit Pauli spin
operators are given in Refs. \citep{singh-pra-16} and
\citep{singh-pra-18}, respectively. As an example, for the
measurement of the moment $ \sigma^{}_{1x} \sigma^{}_{2x}
\sigma^{}_{3x} $,  the form of $U_l=CNOT_{23}.\overline{Y}_3.CNOT_{12}.\overline{Y}_2.\overline{Y}_1$ can be utilize to construct the quantum circuit \citep{singh-pra-18}. Similarly any desired moment can be
measured precisely utilizing these techniques.\\

As stated earlier, the information regarding local/non-local
nature of the observed correlations gets encoded in the
measured correlators $ \lbrace \langle A_x \rangle,\;\langle
B_y \rangle, \;\langle C_z \rangle,\; \langle A_x B_y
\rangle,$ $\langle A_x C_z \rangle,\; \langle B_y C_z
\rangle,\; \langle A_x B_y C_z \rangle \rbrace $. The
formulated SDP in all the cases, \ie W, GHZ and graph
states, failed to find $ \Gamma \geq 0 $ at the second level
of the modified NPA hierarchy. This confirmed that the
observed correlations can not arise from the local
measurements on a separable state and hence the states are
genuinely entangled. A bar plot for the observable moments
of the moment matrix $\Gamma$ for W-state and GHZ-state is
depicted in Fig.\ref{gammaplots}(a) and
Fig.\ref{gammaplots}(b), respectively.
Fig.\ref{gammaplots-grpahstates}(a) and
Fig.\ref{gammaplots-grpahstates}(b) shows the results for
linear and loop graph states respectively. It is interesting
to note from the bar plots of Fig(\ref{gammaplots}) that the
correlations of GHZ state are mostly encoded in the full
three-body correlators while for the W state fewer body
moments are also able to capture the information about
genuine entanglement. Similarly, one can observe in
Fig.\ref{gammaplots-grpahstates} the role of various moments in encoding the information about the correlations for the linear as well as loop graph states. It may be noted here that only experimentally non-vanishing moments are shown in Fig.\ref{gammaplots-grpahstates} for graph states while Fig.\ref{gammaplots} shows all the observed moments in case of W as well as GHZ states.

In all the cases, the SDP was also formulated directly from
experimentally reconstructed density matrices using full
quantum state tomography. This verified and further
supported the results of the modified NPA protocol obtained
via direct measurements of the correlators. We note here in
passing that the experimental protocol demonstration was on
pure states, but the scheme is also capable of detecting
non-locality of states which are a convex sum of white noise
and pure states, up to a certain degree of mixedness
\citep{flavio-prx-17}.

\section{Concluding remarks} 
\label{remarks} 
Local measurement-based hierarchies can be used to detect
the presence of
non-local correlations in a composite quantum system.
A modified NPA hierarchy was used to detect the non-local nature
of the tripartite correlations in a system of three qubits,
by performing local measurements evolved via a semi-definite
program.  A set comprising of products and/or linear
superpositions of such products of local projectors was
defined and an associated positive semi-definite moment
matrix was set up. Non-local correlation detection protocols
typically require the experimental measurement of some
correlator, in order to generate the statistics to be
tested. Once the moment matrix embedded with the empirical
data is obtained, the semi-definite program optimizes to
obtain a positive moment matrix, under some linear
constraints on the entries of the moment matrix, to check if
the observed correlations can arise from local measurements
on a separable state. The protocol was tested
experimentally on three-qubit W, GHZ and a few graph
states utilizing NMR hardware. In all the cases, the
resulting SDP turned out to be unfeasible at the second
level of the modified hierarchy, implying successful
detection of the non-local nature of the observed
correlation. The results were also verified by direct
full quantum state tomography and then directly computing
the correlators to formulate the SDP. Future directions of
research include evaluating the performance of the protocol
in higher dimensions as well as for more number of entangled
parties, since the structure of the entanglement classes is
more complex in a higher-dimensional Hilbert space.

\begin{acknowledgments}
All the experiments were performed on a Bruker Avance-III
600 MHz FT-NMR spectrometer at the NMR Research Facility of
IISER Mohali. 
\end{acknowledgments}

%

\end{document}